\relax
\documentclass[letterpaper]{article} 
\usepackage{aaai22}  
\usepackage{times}  
\usepackage{helvet}  
\usepackage{courier}  
\usepackage[hyphens]{url}  
\usepackage{graphicx} 
\usepackage{natbib}  
\usepackage{caption} 
\DeclareCaptionStyle{ruled}{labelfont=normalfont,labelsep=colon,strut=off} 
\usepackage{algorithm}
\usepackage{algorithmic}
\usepackage[ruled, noline, noend, nofillcomment, algo2e]{algorithm2e}
\usepackage{comment}
\usepackage{amsfonts,amsmath,mathtools}
\usepackage{newfloat}
\usepackage{listings}
\floatstyle{ruled}
\newfloat{listing}{tb}{lst}{}
\floatname{listing}{Listing}
\usepackage[]{xcolor}

\newcommand{\cE}{\mathcal{E}}
\usepackage{xspace}

\newcommand{\ourmaintitle}{Differentially Describing Groups of Graphs} 

\newcommand{\ourmethod}{\textsc{Gragra}\xspace}
\newcommand{\oururl}{\url{https://doi.org/10.5281/zenodo.6342823}}

\newif\ifapx
\apxfalse 

\usepackage{booktabs}

\title{\ourmaintitle}
\author{Corinna Coupette,\!\equalcontrib{}\textsuperscript{\rm 1} Sebastian Dalleiger,\!\equalcontrib{}\textsuperscript{\rm 2} Jilles Vreeken\textsuperscript{\rm 2}}
\affiliations{
	\textsuperscript{\rm 1} Max Planck Institute for Informatics\\
	\textsuperscript{\rm 2} CISPA Helmholtz Center for Information Security\\
	
}

\usepackage{amsmath}
\usepackage{amssymb}
\usepackage{multirow}

\newcommand{\cG}{\mathcal{G}}

\providecommand{\abs}[1]{\left\lvert#1\right\rvert} 
\DeclareMathOperator*{\argmax}{arg\,max}
\DeclareMathOperator*{\argmin}{arg\,min}
\newcommand{\BIC}{\ensuremath{\operatorname{BIC}}}
\newcommand{\Expectation}{\ensuremath{\operatorname{\mathbb{E}}}}

\newcommand{\df}{\ensuremath{\operatorname{df}}}

\begin{document}

\maketitle

\begin{abstract}
How does neural connectivity in autistic children differ from neural connectivity in healthy children or autistic youths? 
What patterns in global trade networks are shared across classes of goods,
and how do these patterns change over time?
Answering questions like these requires us to differentially describe groups of graphs: 
Given a set of graphs and a partition of these graphs into groups, 
discover what graphs in one group have in common, 
how they systematically differ from graphs in other groups, 
and how multiple groups of graphs are related. 
We refer to this task as \emph{graph group analysis}, 
which seeks to describe similarities and differences between graph groups by means of statistically significant subgraphs.
To perform graph group analysis, we introduce \ourmethod, 
which uses maximum entropy modeling to identify a non-redundant set of subgraphs with statistically significant associations to one or more graph groups.
Through an extensive set of experiments on a wide range of synthetic and real-world graph groups, we confirm that \ourmethod works well in practice.

\end{abstract}

\section{Introduction}
\label{sec:intro}

Differentially describing groups of graphs lies at the heart of many scientific and societal challenges.
Neuroscientists, for example, might want to characterize brain activity in healthy subjects, elucidate how it differs from brain activity in subjects diagnosed with certain disorders or diseases (e.g., autism or Alzheimer's), 
and investigate whether their findings are the same across different groups of subjects (e.g., children, adolescents, and adults; or men and women).
Policymakers, security experts, and epidemiologists alike could seek to understand patterns of human mobility, 
be it to improve the resilience of traffic infrastructure to random failures and targeted attacks, 
or to curb the spread of infectious diseases.
And international economists might want to investigate patterns of world trade, e.g., imports and exports between countries, and ask how these vary across different years and product classes.

We refer to the common task underlying these scenarios as \emph{graph group analysis}: 
Given a set of graphs and a partition of this set into \emph{graph groups}, succinctly summarize the commonalities and differences 
between graphs in the same group, 
between graphs in different groups, 
and between the relationships connecting the groups.
In this paper, we formalize graph group analysis as a maximum likelihood modeling problem, 
using \emph{significant subgraphs} as graph patterns to factorize our probability distribution.
We introduce \ourmethod (\emph{Gra}ph \emph{gr}oup \emph{a}nalysis) as an algorithm to solve this problem, 
which jointly discovers a set of graph patterns and an assignment of these patterns to graph groups.

\begin{figure}[t!]
	\centering
	\includegraphics[height=4.5cm]{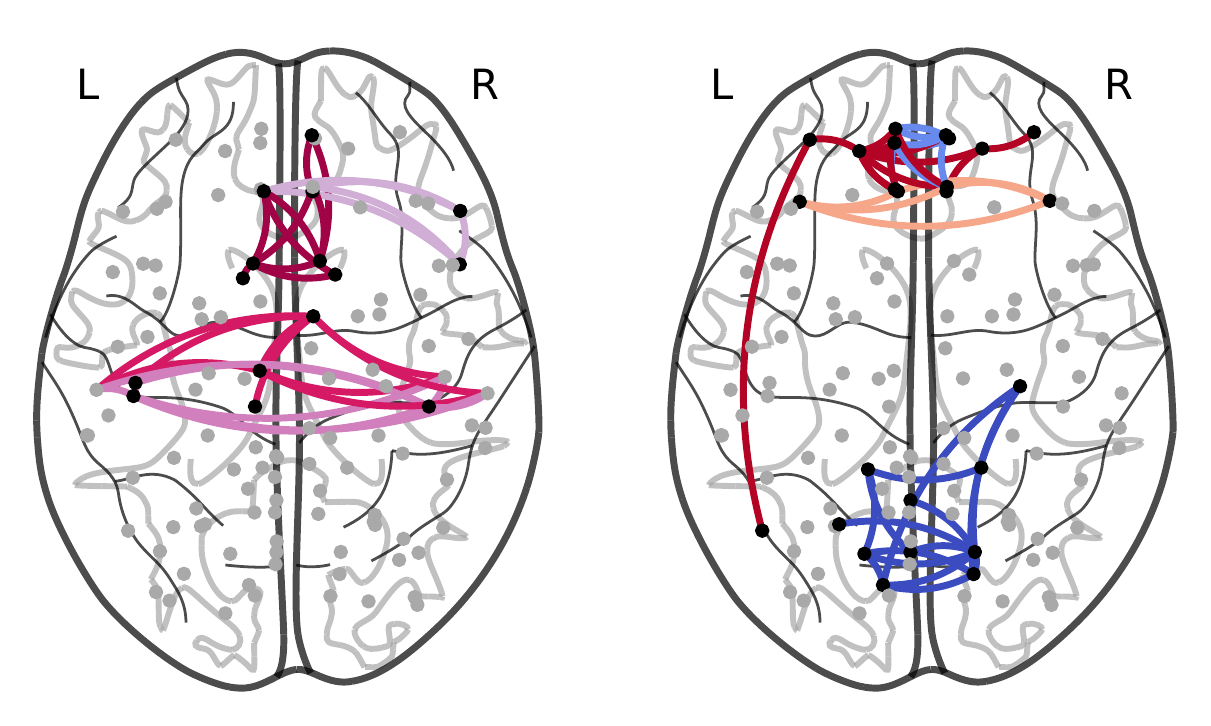}
	\caption{\ourmethod discovers common and contrastive graph patterns in noisy, heterogeneous groups of graphs, 
		capturing, e.g., systematic similarities (left) and differences (right) between the functional brain networks of adolescents with and without autism spectrum disorder. 
		Here, nodes represent centers of mass for brain regions from the AAL Atlas, 
		and edge color classes correspond to significant subgraphs shared between (left) or specific to (right) groups, with individual edges signaling strong connectivity between regions.
	}\label{fig:brains1}
\end{figure}

As a real-world example of graph group analysis, consider Fig.~\ref{fig:brains1}.
Here, we show the top shared (left) and specific (right) patterns identified in resting-state functional brain networks of adolescents with and without autism spectrum disorder, 
where nodes in the graphs correspond to brain regions, 
and edges signal strong connectivity between regions.
On the right, patterns with red edges are characteristic of autistic adolescents, 
and patterns with blue edges are characteristic of non-autistic adolescents.
They indicate over- and underconnectivity, respectively, 
in the brains of autistic adolescents when compared to typically developed controls. 
Although there is no consensus regarding the relationships between autism and neural connectivity 
\cite{hull:17:resting}, 
our method identifies graph patterns that permit neuroscientific interpretation: 
For example, the dark blue pattern in Fig.~\ref{fig:brains1}
indicates underconnectivity between the visual cortex, 
responsible for processing visual information, 
and the lingual gyrus, involved in vision and word processing.

Graph group analysis is related to \emph{graph classification} \cite[e.g.,][]{lee:18:graph}, 
but we are interested not only in what is \emph{different} but also in what is \emph{similar} among our graph groups.
Our task further shares some of its motivation with \emph{significant subgraph mining} \cite[e.g.,][]{sugiyama:15:significant}, 
\emph{graph summarization} \cite[e.g.,][]{liu:18:graph},
and \emph{data clustering} with graphs as data points \cite[e.g.,][]{mukherjee:17:clustering}. 
However, we focus on a \emph{complete} characterization of a \emph{set} of graphs under a \emph{given} partition---%
a cornerstone of scientific discovery involving graph data.

The remainder of the paper is structured as follows. 
After settling our basic notation, we describe the theoretical foundations of our method and introduce our algorithm.
Having covered related work,
through experiments on synthetic and real-world data, 
we demonstrate that \ourmethod works well in practice,
before rounding up with discussion and conclusions.
All our data, code, and results are publicly available.\!\footnote{\oururl}

\section{Preliminaries}
\label{sec:prelim}

We consider a set $\cG = \{G_1,\dots,G_{|\cG|}\}$ of $|\cG|$ \emph{node-aligned} graphs  $G_i=(V,E_i)$ with $n = |V|$ nodes and  $m_i=|E_i|$ edges, 
omitting the subscripts when clear from context.
A partition $\Pi = \{\cG_1, \dots, \cG_k\}$ is a set of $k$ non-empty subsets of $\cG_i\subseteq\cG$, 
called \emph{graph groups}, of cardinalities $c_i = |\cG_i|$, 
whose disjoint union is $\cG$. 
Our graphs can be undirected or directed,  loopy or non-loopy, and unweighted, edge-labeled, or integer weighted, 
where for the purposes of our model, 
we treat distinct edge labels or edge weights as a set $W$ of categories,  
and edges $e\in E_i$ are drawn from the set $\cE = V \times V \times W$ of all possible weighted edges.

The \emph{empirical frequency} of edge set $X\subseteq \cE$ in group $\cG_i$ is 
$q_i(X) = \abs{\{ (V,E) \in \cG_i \mid X \subseteq E \}}/c_i$,
and we denote by $V_X$ the set of nodes incident with at least one edge in $X$. 

We base our probabilistic model on the maximum entropy principle, 
by which the distribution that best reflects a given set of constraints without introducing additional assumptions 
is the distribution with maximum Shannon entropy \cite{jaynes:82:rationale}.
Thus, the expected frequency of $X$ in $\cG_i$ under a given set of edge sets $S\subseteq 2^{\cE}$ is  
\begin{align*}
	p_i(X \mid S) = \Expectation_f[X] = \sum_{\substack{Y \in 2^{\cE},~X \subseteq Y}} f(Y \mid S) \;, 
\end{align*}
where $2^{\cE}$ is the power set of $\cE$, 
and $f$ is the distribution maximizing \(\argmax_f \{-\sum f_X \log f_X\}\), 
subject to linear constraints \(\Expectation_f[X] = q_i(X)\) for all elements in $S$ \cite{csiszar_i-divergence_1975}. 
That is,
\begin{align*}
	f(X \mid S) = \theta_0 \prod_{\substack{Y_i \in S,~Y_i \subseteq X}} \theta_i\;,
\end{align*}  
where $\theta_0$ and all $\theta_i$ are real-valued model parameters.
Finding the distribution $f$ is a convex problem that involves computing the expected frequency $p_i(X \mid S)$ over exponentially many elements. 
This is intractable if done na\"ively, 
but there exist practical approaches that factorize $p_i$ into a product of independent distributions~\cite{mampaey:12:mtv,dalleiger:20:disc,dalleiger:20:reaper}.

\section{Theory}
\label{sec:theory}

We now lay the theoretical foundations of our method, 
introducing our probabilistic model, our objective function, and our statistical test.
At a high level, our goal in graph group analysis is to discover a set $S$ of \emph{graph patterns}, 
i.e., edge sets of \emph{connected subgraphs}, 
and an \emph{association matrix} $A$ assigning graph patterns to graph groups, 
such that $S$ and $A$ together reveal the similarities and differences between graphs in the same group, 
between graphs in different groups, 
and between the relationships connecting the groups.
A pattern is \emph{specific} if we assign it to only one graph group, and it is \emph{shared} if we assign it to several graph groups. 
We choose which patterns to include in our model based on the information we gain from them, 
testing whether this gain is statistically significant to rule out spurious results.

To avoid redundancy, 
we assign a pattern $X$ to a group $\cG_i$ iff $X$ is \emph{informative} for $\cG_i$,
given what we already know about all groups.
More precisely, using $X$ as a column index of $A$ in a slight abuse of notation, 
we set  $A_{iX} = 1$ iff $X$ is \emph{informative} for $\cG_i$ under our current model $(S,A)$.
We assess this by comparing the \emph{empirical frequency} of $X$ in group $\cG_i$,  $q_i(X)$, to its \emph{expected frequency} in that group under our current model,
$p_i(X\mid S_i)$,
where 
$S_i = \{ X \in S \mid A_{iX} = 1 \}$, 
and $p_i$ is obtained from a practical approximation of the maximum entropy distribution with constraint set $S_i$.
$X$ is \emph{informative} for $\cG_i$ iff $q_i(X)$ is significantly different from $p_i(X \mid S_i)$, 
as judged by a statistical test,
and we add $X$ to $S$ (and column $X$ to $A$) 
if $X$ is informative for \emph{some} $\cG_i\in\Pi$.

To identify a suitable set of graph patterns $S$ and an adequate association matrix $A$, 
we exploit the interplay between two steps.
First, we discover the best pattern $X$ to add to $S$, 
given the current $(S,A)$,
and second, we identify the best assignment of $X$ to graph groups to update $A$, 
given the current $(S,A)$ \emph{and} a new pattern $X$.
We now describe each step in more detail.

\subsection{Identifying Informative Graph Patterns} 

To measure the likelihood of a set $S\subseteq 2^{\cE}$ of graph patterns, 
we use the \emph{Bayesian Information Criterion} (BIC),
\begin{align*}
    \BIC(S) = \ell(S) + (k\cdot\abs{S})/2 \log |\cG|\quad\text{\cite{schwarz:78:estimating}},
\end{align*}
where $k\cdot\abs{S}$ is the number of coefficients in our model, and
\begin{align*}
	\ell(S) = \sum_i \ell_i(S) = -\sum_i \sum_{G \in \cG_i} \log p_i(G \mid S_i)\,,
\end{align*}
is the log likelihood of $S$ (with $S_i \subseteq S$ derived from $A$), assuming that the graphs in a group are independent and identically distributed.
This allows us to identify a good set of graph patterns by minimizing the BIC score, i.e.,
\begin{align*}
    \argmin_{S \subseteq 2^{\cE}}\ \{\BIC(S)\}\;. 
\end{align*}
Solving this problem exactly poses significant challenges in practice 
due to its combinatorial nature and the explosion in the number of solution candidates.
Therefore, we employ a greedy search strategy, 
iteratively selecting the graph pattern $X \subseteq \cE$ that best improves our current model. 
That is, for a given $(S,A)$, 
we select the graph pattern $X$ that maximizes our likelihood, or equivalently, maximizes the difference \(\BIC(S) - \BIC(S \cup \{X\})\), which we write as 
\begin{align*}
    \Delta(X) = \ell(S) - \ell(S \cup \{X\}) - k/2\log |\cG|\;.
\end{align*} 
In a nutshell, the core of our approach is the procedure
\begin{equation}
    S \gets S\; \cup \big\{\argmax_{X \subseteq \cE,~\Delta(X) > 0}\ \{\Delta(X)\}\big\}\;,
    \label{eq:greedy}
\end{equation} 
by which we iteratively and greedily insert into $S$ the pattern $X\subseteq\cE$ that locally maximizes our information gain.

Using a model selection criterion alone, however, 
we cannot tell if our information gain is due to random fluctuations or due to signal, 
especially if we only have a limited number of samples. 
Thus, to avoid modeling noise, 
we add $X$ to $S$ only if its information gain $\Delta(X)$ is \emph{statistically significant}.
Therefore, we test whether we can reject the null hypothesis 
\begin{align*}
	H_0:\; \BIC(S) = \BIC(S \cup \{X\})\;.
\end{align*}
To this end, we use Vuong's closeness test~\cite{vuong:89:likelihood}, 
a likelihood ratio test designed for model selection problems under BIC.
Vuong's test statistic is defined as $2\Delta(X)$, 
which is asymptotically $\chi^2$-distributed with 
$\df_{\Delta}(X)=\df p_i(\ \cdot \mid S \cup \{X\}) - \df  p_i(\ \cdot \mid S )$
degrees of freedom.
To calculate $\df_{\Delta}(X)$, 
we count the coefficients $\theta$ that must be changed in every distribution if we insert $X$ into $S$. 
As we add one coefficient for $X$, 
and update at least $|X|$ edge coefficients per group, 
we arrive at $|X|+1$ additional degrees of freedom.

\subsection{Discovering Differential Pattern Associations} 

Once we have selected a new pattern $X\subseteq \cE$ to add to $S$, given the current $S$ and $A$, 
we identify a good assignment of $X$ to graph groups $\cG_i\in\Pi$ to update $A$.
Here, the significance of $\Delta(X)$, which is used to accept $X$ into $S$,
signals that $X$ is informative for \emph{some} $\cG_i\in\Pi$, 
but it does not tell us for \emph{which} $\cG_i$.
To assign $X$ to a graph group $\cG_i$, we hence rely on the \emph{partial} information gain of $X$ for $\cG_i$,
\begin{align*}
    \Delta_i(X) = \ell_i(S_i) - \ell_i(S_i \cup \{X\}) - k/2\log |\cG|\;.
\end{align*}
Again, we use Vuong's closeness test to decide whether $\Delta_i(X)$ is significant; 
and if it is, we set $A_{iX} = 1$.

\section{Algorithm}
\label{sec:algo}

Having stated its theoretical foundations, 
we now introduce \ourmethod as an algorithm to differentially describe groups of graphs using sets of significant subgraphs.
\ourmethod, 
whose pseudocode is given as Alg.~\ref{alg:gragra},
revolves around the procedure stated in Eq.~\eqref{eq:greedy}, 
a greedy process that iteratively selects the graph pattern candidate that best enhances our model.
This \emph{could} involve myriad searches through the exponentially-sized space of all possible graphs with nodes from $V$,
which is not only computationally infeasible in most cases 
but also unnecessary, 
as most candidates will be eliminated as uninformative by Vuong's test. 
Hence, rather than exhaustively searching for the best graph patterns,
we propose to grow graphs by systematically adding edges to candidates.

\begin{algorithm2e}[t]
	\DontPrintSemicolon
	\LinesNumbered
	\SetKwIF{If}{ElseIf}{Else}{if}{}{else if}{else}{end if}%
	\SetKwFor{While}{while}{}{end while}%
	\SetKwFor{For}{for}{}{end while}%
	\SetNlSty{tiny}{}{}%
	\SetKwProg{Fn}{Fn.}{ }{end}
	\SetKwProg{Alg}{Alg.}{ }{end}
	\SetKwRepeat{Do}{do}{while}
	\KwIn{groups of graphs $\cG_1,\dots,\cG_k$}
	\KwOut{set of graph patterns $S$, association matrix $A$}
	
	$S \gets \cE$\;
	$A \gets$ empty binary matrix with $k$ rows and $0$ columns\; 
	$C \gets \{ \{x,y\} \mid x,y \in \cE, x\neq y, V_{\{x\}} \cap V_{\{y\}} \neq\emptyset \}$ \;\label{l:candidates}
	\While{$C \neq \emptyset$}{
		$\hat{X}, C \gets \textsc{Grow}(C)$ \;
		\If{$\exists i \in [k]$ \emph{s.t.} $h_i(\hat{X})$ \rm is significant}
		{
			resize $A$\;
			$A_{i\hat{X}} = 1 \iff h_i(\hat{X})\;\text{is significant}\;\forall i \in [k] $\;\label{l:groupassoc}
			$S \gets S \cup \{\hat{X}\}$\;\label{l:insert}
			estimate $p_i(\ \cdot\mid S_i)\;\forall i \in [k]$ s.t. $A_{i\hat{X}} = 1$\;
		}
	}
	\Return{$S\setminus \cE, A$}\;
	\;
	\Fn(\label{l:grow:beginning}){\textsc{Grow}$(C)$}{
		$\displaystyle X \gets \argmax_{X \in C}\; \{h(X)\;\text{s.t.}\;h(X)\;\text{is significant}\}$\;\label{l:select}
		
		$\displaystyle C \gets C\; \cup\; (((V_X \tiny{\times} V \tiny{\times} W) \;\cup\; (V \tiny{\times} V_X \tiny{\times} W))\setminus X)$\;\label{l:grow}
		$C \gets \{ X \in C \mid h(X)\ \text{is significant}\ \}$\;\label{l:significant}
		$\displaystyle\hat{X} \gets \argmax_{X \in C} \{h(X)\}$\;
		\eIf(\label{l:fullpotentialcond}){$h(\hat{X}) > h(X)$}{
			\Return{$\textsc{Grow}(C)$}\label{l:fullpotential}
		}
		{
			\Return{$\hat{X}, C \setminus \{\hat{X}\}$}\label{l:grow:end}
		}
	}
	\caption{\textbf{\ourmethod}} 
	\label{alg:gragra}
\end{algorithm2e}

To enable our model to infer all possible graphs, we initialize it with the set $\cE$ of all possible edges.
As our initial graph to grow, we then select the most promising graph pattern from our initial candidates, i.e., the connected triples
\begin{align*}
	C = \{ \{x,y\} \mid x,y \in \cE, x\neq y, V_{\{x\}} \cap V_{\{y\}} \not=\emptyset \}\;.
\end{align*} 
Starting with a graph pattern $X$, 
we explore all its expansions, $((V_X \times V \times W) \cup (V \times V_X \times W)) \setminus X$, 
from which we select the best candidate pattern to grow further, 
as long as we gain information and $\Delta(X)$ is significant.
We summarize these steps in the function~\textsc{Grow} of Alg.~\ref{alg:gragra} (l.~\ref{l:grow:beginning}--\ref{l:grow:end}).

\textsc{Grow} requires many inferences of $\Delta$, 
which involve inferring many more expected frequencies $p_i$, 
rendering exact computation impractical.
We thus design a practical, pessimistic heuristic that only considers the information gain from graphs $G \in \cG$ in which $X$ is fully present:
\begin{equation}
    h(X) = \sum_i c_i \cdot q_i(X) \log \frac{q_i(X)}{p_i(X)} - k/2 \log |\cG|\;.
	\label{eq:heuristic}
\end{equation}
We use $h(X)$ instead of $\Delta(X)$ 
because it involves inferring only \emph{one} expected frequency per graph group. 
A derivation of this heuristic can be found in the Online Appendix.

To summarize, \ourmethod proceeds as follows. 
Starting with an initial set of candidates $C$ (l.~\ref{l:candidates}),
we select (l.~\ref{l:select}) and grow (l.~\ref{l:grow}) the best candidate,
and retain all significant expansions (l.~\ref{l:significant}),
until we have grown $X$ to its fullest potential (l.~\ref{l:fullpotentialcond}--\ref{l:fullpotential}).
Afterwards, we test if the information gain provided by $X$ is significant, 
and if so, we keep track of its graph group associations (l.~\ref{l:groupassoc}),
and insert $X$ into $S$ (l.~\ref{l:insert}).

The computational complexity of \ourmethod depends on the number of candidates, which can grow to at most $|2^{\cE}|$. 
In practice, \ourmethod's complexity depends on the number of times we grow graph patterns, 
which is data-dependent and bounded by the size $\gamma$ of the largest connected component observed in an input graph, 
as growing beyond that reduces the information gain.
Multiplying $\gamma$ by the initial set of candidates, 
\ourmethod achieves a complexity of $\mathcal{O}\big(\binom{n}{3}|W|\gamma\big)$ for all practical purposes, 
where we assume that the complexity of inferring the expected frequency is bounded.

\section{Related Work}
\label{sec:related}

To the best of our knowledge, we are the first to differentially describe groups of graphs through sets of significant subgraphs. 
Our method is inspired by advances in graph similarity description \cite[\textsc{Momo},][]{coupette:21:graph}
and explainable pattern set mining using maximum-entropy modeling \cite[\textsc{DISC},][]{dalleiger:20:disc,dalleiger:20:reaper}. 
However, \textsc{Momo} focuses on pairs and unpartitioned sets of graphs; 
\textsc{DISC} is designed for itemset data, ignores graph structure, and does not scale on graphs; 
and neither method uses a statistical test to select patterns.
Further related work broadly falls into two categories: 
statistical inference on network populations, and graph mining for groups of graphs. 

\emph{Statistical Inference on Network Populations.}
In the statistics literature, the task of analyzing multiple graphs simultaneously is typically framed as an inference problem for network-valued random variables \cite{durante:17:nonparametric,lovato:20:model,lunagomez:20:modeling}.
Here, \citet{ghoshdastidar:20:hypothesis} establish limits for distinguishing two population distributions given small sample sizes, 
and \citet{lunagomez:20:modeling} propose notions of mean and dispersion for a single population of networks, where the population mean is itself a network.
\citet{maugis:20:testing} use subgraph counts to test if all graphs in a sample are drawn from the same distribution,
and \citet{signorelli:20:model} propose a model-based clustering approach to describe subpopulations within a population of networks.  
Finally, \citet{durante:17:nonparametric} extend latent space approaches designed for single graphs 
to capture the probabilistic mechanism that generates multiple graphs from a single population distribution.
Their model has been used to characterize and test for differences between groups of brain networks \cite{durante:18:bayesian}---%
an actively studied application for which numerous statistical methods, 
mostly focusing on \emph{testing} for differences, 
have been developed \cite{ginestet:17:hypothesis,wang:19:common,lukemire:20:bayesian,kundu:21:learning,lovato:21:multiscale,lehmann:21:characterising}.

Prior work in the statistics literature has focused on describing \emph{one} network population or distinguishing \emph{two} populations. 
In contrast, with \ourmethod, 
we aim to construct a \emph{differential description} of \emph{any} number of populations. 
Furthermore, we ask not only \emph{if} these populations are different, 
but also \emph{how} they are different and how they are \emph{similar}.

\emph{Graph Mining for Groups of Graphs.}
In the graph mining literature, groups of graphs have been studied in contexts as diverse as significant subgraph mining \cite{llinares:15:fast,sugiyama:15:significant}, 
graph classification \cite{vogelstein:13:graph,yan:19:groupinn,lanciano:20:explainable}, 
graph clustering with graphs as data points \cite{mukherjee:17:clustering}, 
anomalous graph detection \cite{gomes:18:multi},
and graph summarization for time series of graphs \cite{shah:15:timecrunch}.
Significant subgraph mining commonly considers small, node-labeled graphs with unaligned node sets, and hence, does not target our problem.
However, our setup---i.e., medium-sized graphs with aligned node sets---, has received heightened attention in the graph classification community, again inspired by challenges from neuroscience \cite{vogelstein:13:graph,yan:19:groupinn,lanciano:20:explainable}.

The methods that are closest to our work are \emph{contrast subgraphs} \cite{lanciano:20:explainable} and \emph{signal subgraphs} \cite{vogelstein:13:graph}, 
both designed for two groups of node-aligned graphs.
\emph{Contrast subgraphs} discover the densest subgraph in the difference of the summary graphs of the input groups (obtained by adding the graphs in each group separately and then subtracting the results), 
where the size of this subgraph depends on a user-specified regularization parameter $\alpha$. 
\emph{Signal subgraphs} assume edge independence as a prior to rank edges by the $p$-values of an edge-wise statistical test for distributional difference (e.g., Fisher's exact test).
Like \emph{signal subgraphs}, \ourmethod combines ideas from structural and statistical pattern mining to produce interpretable results that---%
unlike \emph{contrast subgraphs}---are based on a statistical foundation. 
\ourmethod is more exploratory and more flexible than both competitors, however, 
because it treats graph group description as an end in itself and can handle any number of graph groups.

\section{Experiments}
\label{sec:exps}

We now present an extensive evaluation of our algorithm.
To this end, we implement \ourmethod in C++ and expose a Python interface to facilitate experimentation. 
We run our experiments on Intel E5-2643 CPUs with $128$ or $256$~GB RAM, 
testing at a conservative significance level of $1\times 10^{-7}$ (or $1\times 10^{-5}$ when operating with less than $50$ samples),
and make all data, code, and results publicly available.\!\footnote{\oururl}
Our experiments revolve around two questions:
\begin{enumerate}
	\item Can \ourmethod reliably recover the ground truth from groups of synthetic graphs?
	\item Does \ourmethod discover meaningful patterns in groups of real graphs?
\end{enumerate}

\subsection{Recovering Ground Truth from Synthetic Graphs}

To assess the reliability of \ourmethod, 
we run it on groups of synthetic graphs with planted patterns. 
We consider three scenarios, namely,
\begin{enumerate}
	\item summarizing \emph{one} group of graphs, 
	\item differentially describing \emph{two} groups of graphs, and
	\item differentially describing \emph{four} groups of graphs.
\end{enumerate}
In all three scenarios, each graph group consists of $100$ graphs with $100$ nodes, 
and our configurations differ in their planted patterns (type, prevalence, and position) and noise levels. 
A detailed overview of our synthetic data configurations can be found in the Online Appendix.

For each scenario, we report the distribution of precision, recall, and F1 score, computed separately for each group of graphs, 
for the edges of the planted patterns across $100$ graph group datasets sampled with different seeds.
In all scenarios, we compare \ourmethod, which uses BIC with Vuong's closeness test for pattern selection, 
with a variant using only BIC and no statistical test to select patterns ($\ourmethod_{\text{BIC}}$).
For configurations in the second scenario, 
we also compare our results with those from \emph{contrast subgraphs} (CSG) and \emph{signal subgraphs} (SSG), 
described in the previous section. 

\begin{figure}[t!]
	\centering
	\includegraphics[height=3.3cm]{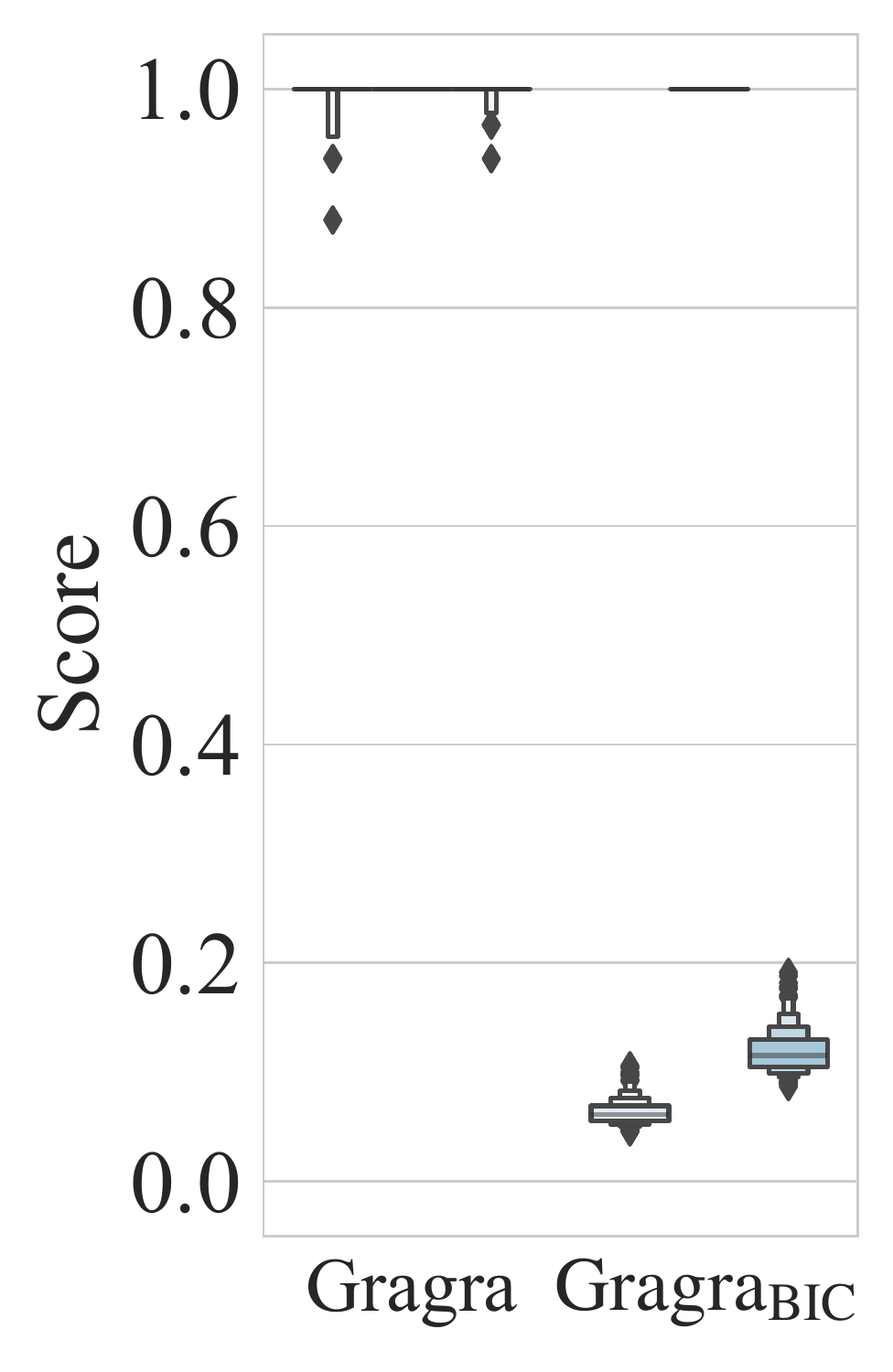}~\hspace*{-0.4em}\includegraphics[height=3.3cm]{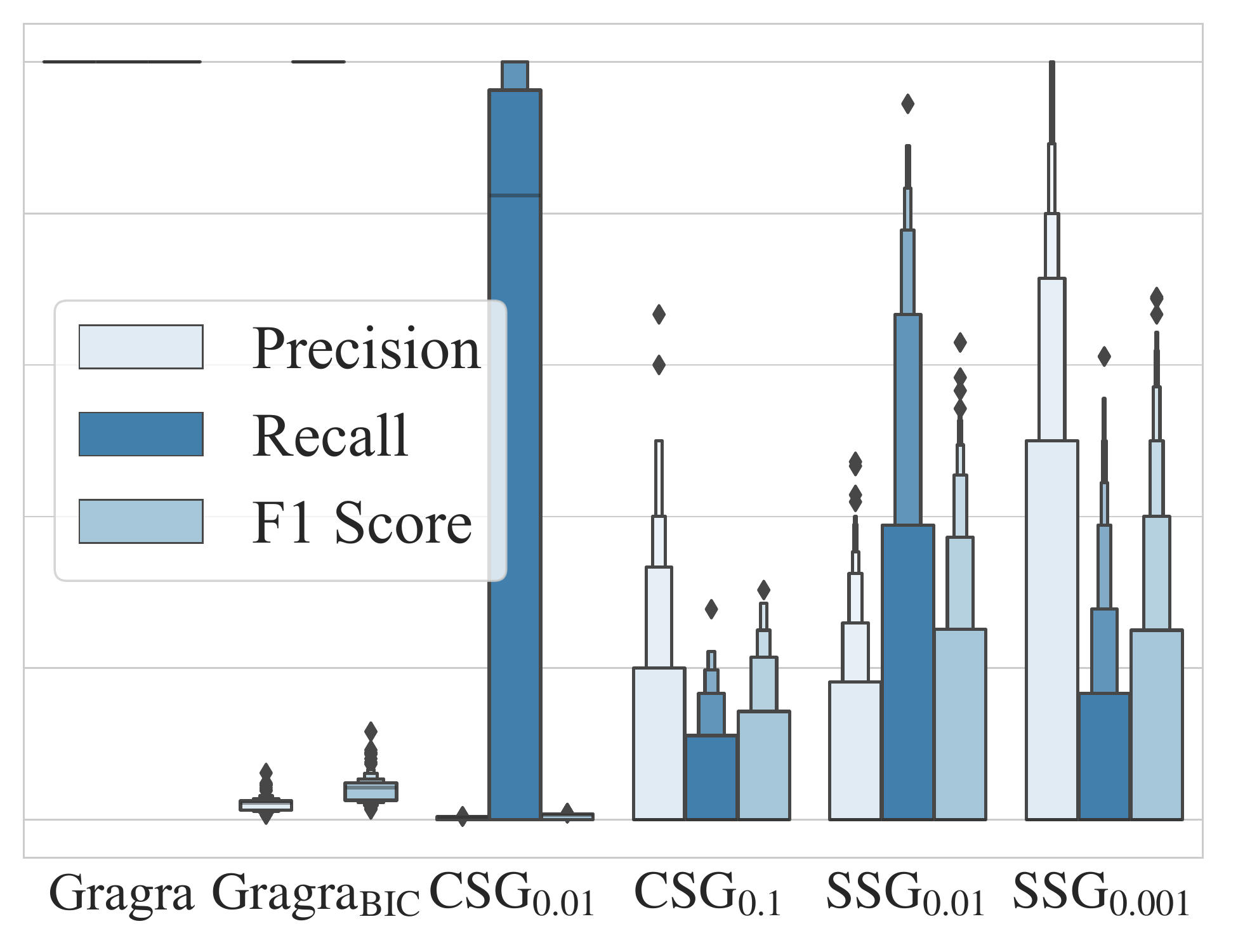}~\hspace*{-0.5em}\includegraphics[height=3.3cm]{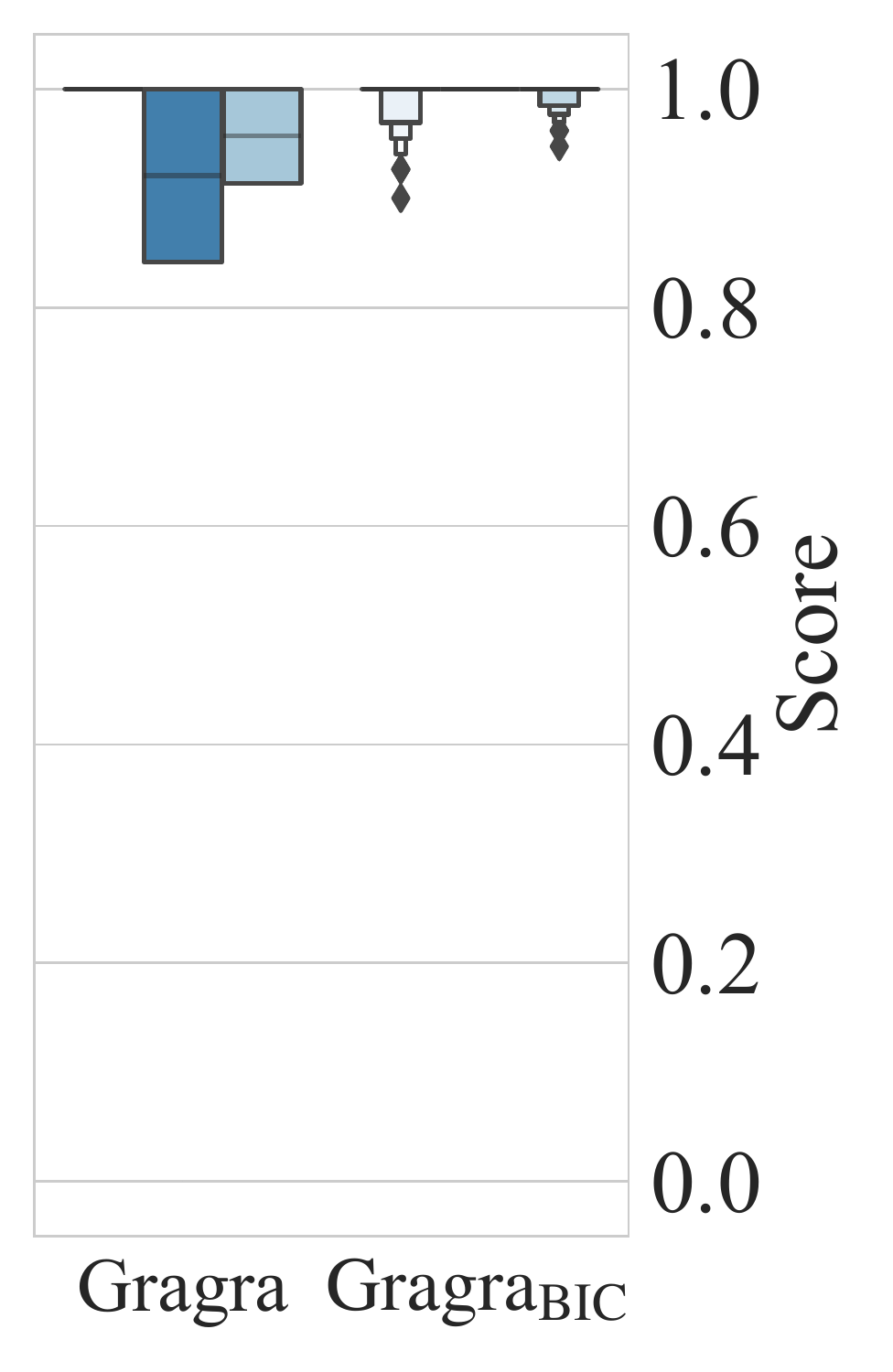}
	\caption{%
		\ourmethod reliably recovers the ground truth from synthetic data.
		We show precision, recall, and F1 score distributions for \ourmethod, 
		$\ourmethod_{\text{BIC}}$,
		\emph{contrast subgraphs} (CSG), 
		and \emph{signal subgraphs} (SSG), 
		separately for all experiments in our three different settings: 
		one-group setting (left), 
		two-group setting (middle), 
		and four-group setting (right). 
		Subscripts of CSG labels correspond to different choices of their regularization parameter $\alpha$, 
		and subscripts of SSG labels indicate different requirements for the $p$-values obtained from their edge-wise distributional difference test.
	}\label{fig:synthetic}
\end{figure}

As shown in Fig.~\ref{fig:synthetic}, 
$\ourmethod_{\text{BIC}}$ delivers good results in the four-group scenario but generally has poor precision, treating noise as signal.
CSG and SSG identify only constrastive patterns, 
and fail even for contrastive patterns 
if the individual edges in planted patterns have similar occurrence probabilities across groups. 
\ourmethod, however, reliably recovers the ground truth across scenarios and configurations, 
which allows us to hope that it will also work in practice.

\subsection{Discovering Meaningful Patterns in Real Graphs}

\begin{figure*}
	\centering
	\includegraphics[height=4cm]{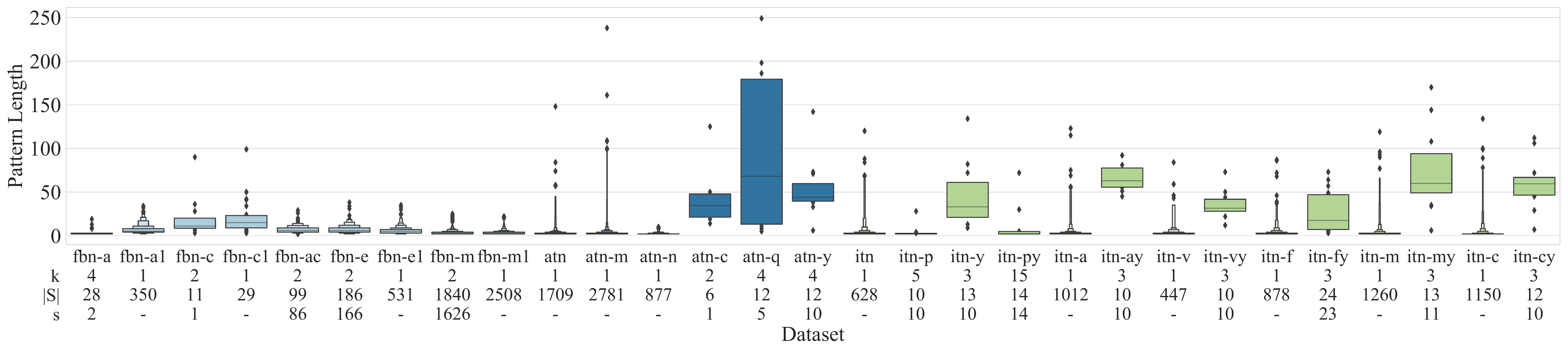}
	\caption{%
		\ourmethod discovers long graph patterns in datasets with different numbers of graph groups. 
		Here, we show the length distribution of the patterns identified in each of our experiments on real-world data, 
		where each box corresponds to a dataset.
		The first number below a dataset identifier states the number of graph groups $k$ in the dataset, 
		the second number states the total number of patterns $|S|$, 
		and the third number states the number of patterns $s$ shared between at least two graph groups.
	}\label{fig:results}
\end{figure*}

To determine whether \ourmethod discovers meaningful patterns in groups of real graphs, 
we run $29$ experiments on graph group data of various graph types from three domains:
functional brain networks (undirected, unweighted), 
air transportation networks (directed, weighted), 
and international trade networks (directed, weighted).
We compile basic statistics of these networks in the Technical Appendix, 
and present a quantitative overview of our results in Fig.~\ref{fig:results}.
We observe that, in line with expectations derived from theory, 
\emph{more graphs} or \emph{graphs with more potential edges}, 
partitioned into \emph{fewer groups}, generally yield \emph{more patterns}. 

\subsubsection{Functional Brain Networks}

\emph{Network neuroscience} has emerged as a promising approach to understanding neurological disorders and diseases \cite{bullmore:09:complex,bassett:17:network,fornito:15:connectomics}.
One of its fundamental questions is 
whether certain disorders are systematically associated with structural or functional connectivity alterations in the brain \cite{van:19:cross}.
In particular, 
there is considerable uncertainty surrounding the neurological footprint of autism (and the delineation of its subtypes), 
and small sample sizes as well as covariates 
make many published findings hard to replicate \cite{he:20:nonreplication,king:19:generalizability}.
This calls for methods that can detect signal in the presence of considerable noise and heterogeneity, 
identifying connectivity patterns that are statistically significantly associated with one or more groups of brain networks. 

Motivated by this application, we obtain graphs from preprocessed functional connectomes provided by the Autism Brain Imaging Data Exchange \cite{craddock:13:neuro}. 
In these graphs, each node corresponds to one of the $116$ \emph{regions of interest} from the automated anatomical labeling atlas \cite[AAL,][]{rolls:20:automated}, 
and each edge indicates relatively strong connectivity between two regions, 
as measured by their blood-oxygen-level dependent signal correlation during resting-state functional magnetic resonance imaging.
To facilitate comparisons, the data is processed and grouped as described by \citet{lanciano:20:explainable}, 
but we remove the self-loops (corresponding to perfect self-correlations) that are present in their data. 

\begin{figure}[t!]
	\includegraphics[height=3.3cm]{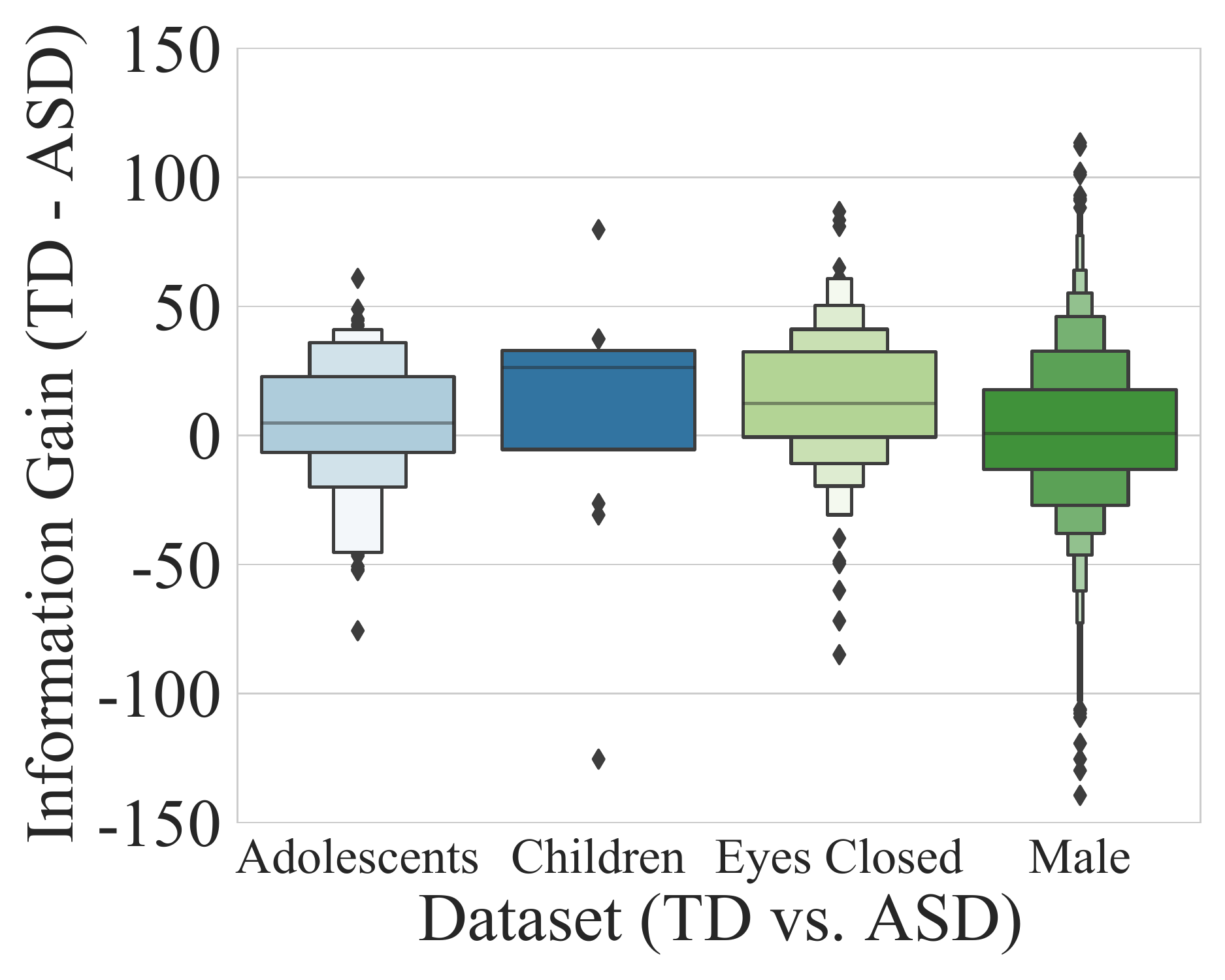}~\includegraphics[height=3.3cm]{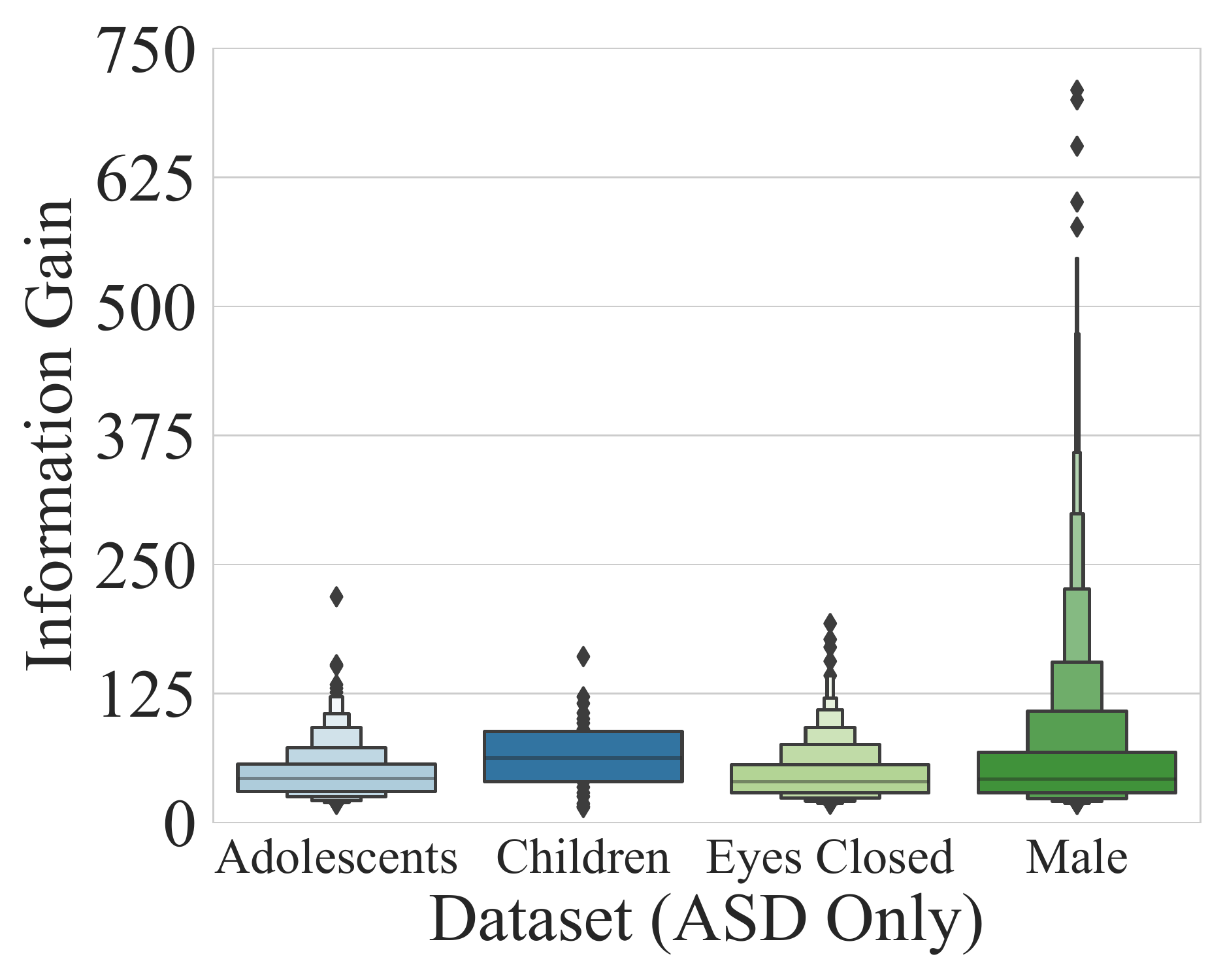}
	\caption{%
		\ourmethod unveils shared and contrastive patterns in noisy and heterogeneous data.
		Here, we display the distribution of information gain differences per pattern in the two-group setting (left), 
		and the distribution of information gains per pattern in the one-group setting (right), 
		for our experiments on functional brain networks.
	}\label{fig:brains2}
\end{figure}

We experiment with \ourmethod in four two-group settings (individuals with autism spectrum disorder [ASD] and typically developed controls [TD] in the categories \emph{adolescents}, \emph{children}, \emph{eyes closed during scan}, and \emph{males}), 
four one-group settings (autistic individuals in each category only), 
and one four-group setting (autistic and non-autistic children and adolescents),
operating on graphs with $m\in[1\,320,1\,348]$ edges and graph groups $\cG_i$ with $c_i\in[49,420]$ graphs.
Our four-group experiment identifies significant overconnectivity across multiple brain regions as characteristic of ASD children versus all other groups, 
paralleling the neuroscience literature \cite{supekar:13:brain,nomi:15:developmental}.
However, as shown in Fig.~\ref{fig:brains2}, 
most of the patterns we identify in the two-group setting yield similar information gains across both groups (left), 
and there is significant structure to be exploited even \emph{within} individual groups (right).
This indicates that the differences between autistic and non-autistic brains in the settings under study are rather subtle, 
and that there is considerable heterogeneity also in the one-group data. 
To explore this heterogeneity and delineate neurosubtypes of autism \cite[cf.][]{hong:20:neurosubtypes}, 
our results could be used as inputs to multivariate subgroup discovery or clustering algorithms, 
where \ourmethod would effectively serve as a dimensionality reduction technique.

\begin{figure*}[t!]
	\centering
	\includegraphics[width=0.37\textwidth]{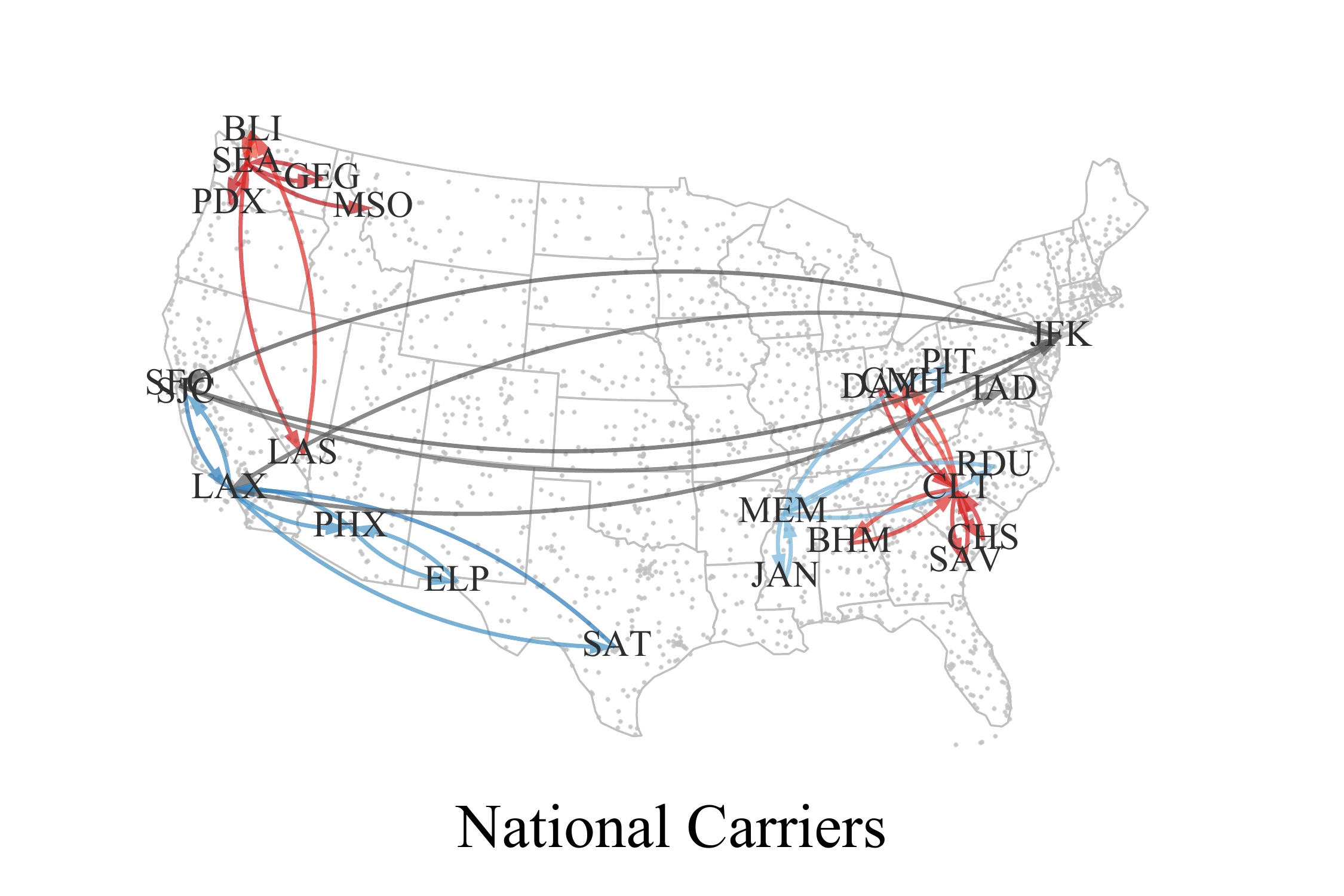}~\hspace*{-2.5em}\includegraphics[width=0.37\textwidth]{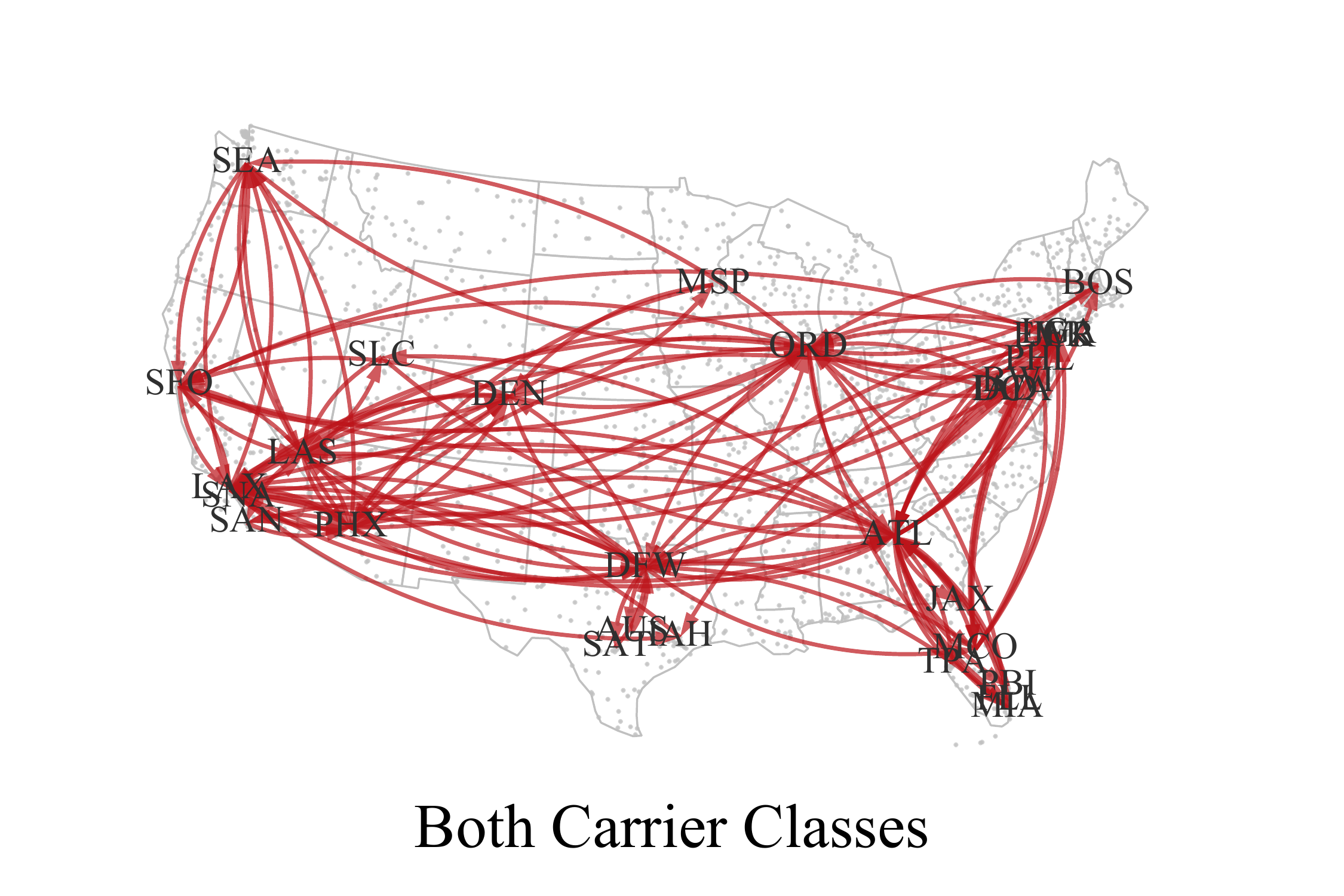}~\hspace*{-2.5em}\includegraphics[width=0.37\textwidth]{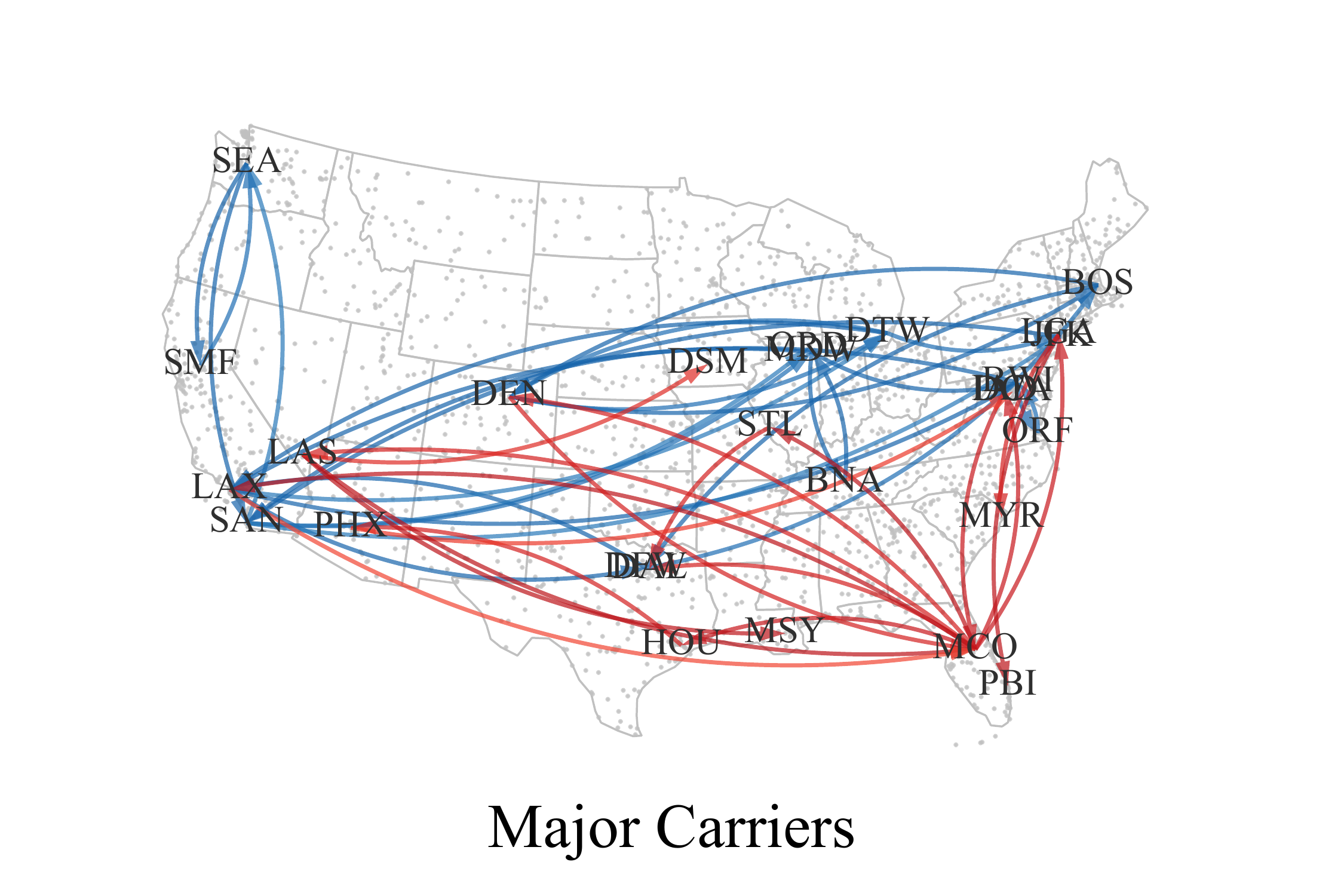}
	\caption{%
		\ourmethod discovers large, meaningful graph patterns.
		Here, we depict some of the patterns discovered in the air transportation networks of national carriers (left, five patterns shown), 
		major carriers (right, two patterns shown), 
		and both carrier classes (middle, one pattern shown).
		Gray nodes represent airports, 
		and node labels identify airports contained in at least one of the displayed patterns by their three-letter IATA codes.
		Directed edges represent flight segments, 
		and edge colors are proportional to their weight bins, 
		following different color maps (reds, blues, or grays) where necessary to make them visually distinguishable. 
		All drawn patterns are among the top fifteen in terms of information gain for their respective experiment, 
		and the pattern in the middle is the top shared pattern,
		corresponding to the United States air transportation backbone.
	}\label{fig:airlines}
\end{figure*}

\begin{figure*}[t!]
	\centering
	\includegraphics[height=3.5cm]{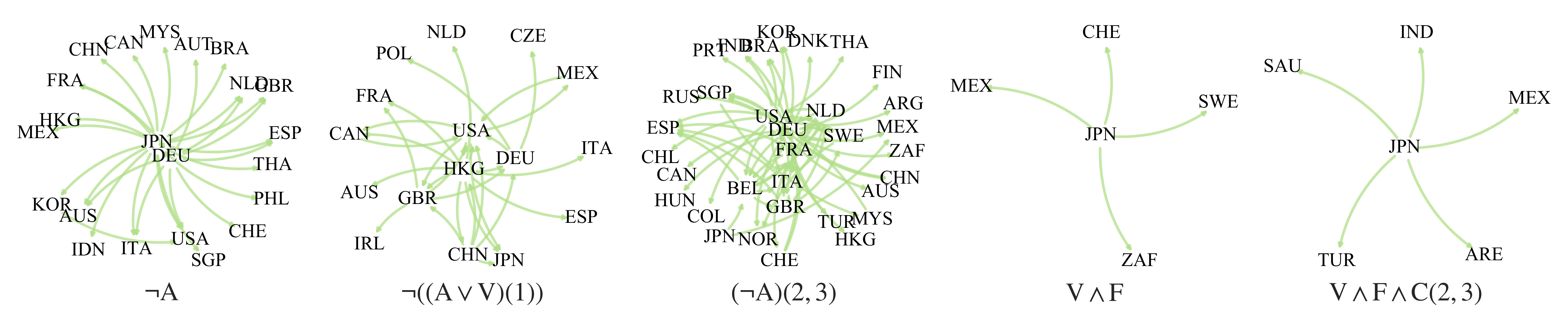}
	\caption{%
		\ourmethod mines differential descriptions even when many graph groups are given as input.
		Here, we show the top five graph patterns identified in the international trade networks when split by product class and decade (fifteen graph groups in total). 
		Nodes correspond to countries, which are represented by their ISO3 country codes. 
		Directed edges correspond to trade flows between the countries, 
		where the edge weights in all displayed patterns fall into the top weight bin.
		The patterns are labeled by rules identifying the graph groups in which they occur, 
		with letters corresponding to the first letter of a product group, 
		and numbers corresponding to the position of a ten-year interval. 
		For example, the third pattern, labeled $(\neg A)(2,3)$, occurs in all product classes except for \emph{Animals}, 
		in the second and the third ten-year interval, i.e., in $[99,19)$.
	}\label{fig:trade}
\end{figure*}

\subsubsection{Air Transportation Networks}

We obtain data on passenger flows between domestic airports in the United States
for each month over the sixteen years from January $2005$ to December $2020$ from the website of the Bureau of Transportation Statistics \cite{bureau:21:airlines}.
Restricting our analysis to United States mainland airports and carriers classified as national ($100$ million to $1$ billion USD revenue in the previous year) or major (over $1$ billion USD revenue in the previous year), 
we create one air transportation network per year, month, and carrier class.
To this end, for each year and month, we aggregate the passenger flows between two airports by carrier class and filter edges corresponding to fewer than $3\,000$ passengers, 
which leaves edges between $n=300$ airports (identified by three-letter IATA codes).
Excluding graphs with fewer than $n-1=299$ edges, we arrive at $374$ graphs, 
whose edges we discretize into ten weight categories using equal-width binning.

We are interested in discovering patterns that are shared across all graphs, 
identifying structures of connected routes that are specific to individual carrier classes, 
and unveiling both seasonal and temporal trends.
Therefore, we run \ourmethod in six different settings: 
on all graphs as one group, 
on the graphs corresponding to each carrier class separately, 
on all graphs with carrier classes as groups, 
on all graphs with quarters as groups (starting from December to capture the winter holiday season), 
and on all graphs with consecutive four-year intervals as groups. 
Thus, our setup contains graphs with $m\in[335,3\,533]$ edges and graph groups $\cG_i$ with $c_i\in[86,374]$ graphs.
In Fig.~\ref{fig:airlines}, we depict a subset of our results from the experiments involving the distinction between carrier classes. 
\ourmethod reveals an air transportation backbone jointly serviced by both carrier classes (middle), 
and it uncovers routes that are characteristically served by national or major carriers (left and right). 
Overall, we find that patterns corresponding to national carrier routes are often smaller and cover shorter distances than those corresponding to major carrier routes, 
mirroring the relatively smaller role of national carriers in the air traffic market.

\subsubsection{International Trade Networks}

We obtain data on international trade flows from the website of the World Integrated Trade Solution \cite{worldbank:21:wits}, 
for the thirty years from $1989$ to $2018$ (inclusive). 
The raw data correspond to exports of goods between (mostly) countries, 
classified using the Harmonized System at the four-digit level (HS-4),
whose trade values we aggregate per (source, destination, HS-4 code) triple.
For each year and HS-4 code, we construct one directed, weighted graph with (roughly) countries as nodes and exports as edges, 
discretizing the edge weights into ten categories using equal-width binning.
We eliminate all trade entities above the country level but retain trade entities below the country level (and countries that do not exist anymore) if they have an ISO3 code. 
Restricting our attention to the WITS product groups \emph{Animals}, \emph{Vegetables}, \emph{Food Products}, \emph{Minerals}, and \emph{Chemicals}, 
we arrive at $3\,976$ graphs with $n=250$ nodes and at least $n-1=249$ edges.

Leveraging the richness of our data, we ask not only what graph patterns are characteristic of international trade as a whole, 
but also what structures emerge when we group trade networks by product class, ten-year interval, or product class \emph{and} ten-year interval. 
As \ourmethod allows us to inspect our data at different scales, 
we further investigate the trade patterns it unveils when considering each product class separately, 
either treating all graphs from one product class as one group 
or splitting them by ten-year interval.
Thus, we run our experiments on graphs with $m\in[256,11\,415]$ edges and graph groups $\cG_i$ with $c_i\in[70,3\,976]$ graphs.
In Fig.~\ref{fig:trade}, we illustrate five patterns discovered in the experiments that explore all graphs together, 
grouped by product class and ten-year interval.
Although the input consists of fifteen classes, 
\ourmethod discovers not only meaningful patterns 
but meaningful patterns \emph{with meaningful assignments} to graph groups which,
as highlighted by the pattern labels in Fig.~\ref{fig:trade}, 
can be summarized succinctly.
Across all experiments, we observe that the patterns yielding the largest information gains are often composed entirely of edges in the top two weight bins.
This suggests that the ranking of exporter-importer pairs is most stable on the upper end of the trade-value spectrum, 
which aligns with interdisciplinary research findings that international trade is highly stratified \cite{sacks:01:global,lloyd:09:looking,fagiolo:10:evolution}.

\section{Conclusion}
\label{sec:conclusion}

We study the \emph{graph group analysis} problem: 
Given a set of graphs and a partition of this set into \emph{graph groups}, succinctly summarize the commonalities and differences 
between graphs in the same group, 
between graphs in different groups, 
and between the relationships connecting the groups.
We introduce \ourmethod as an algorithm to solve the problem, 
which uses maximum likelihood modeling, 
paired with a model selection criterion and a statistical test, 
to jointly discover a set of significant subgraphs, called graph patterns, 
and an assignment of these patterns to graph groups.
In our experiments, 
we demonstrate that \ourmethod differentially describes synthetic and real-world graph groups, 
even when faced with heterogeneity, noise, or large group numbers. 
As a byproduct, we introduce two novel datasets of node-aligned graphs, 
which might be of independent interest to the graph mining community.

However, our work also has limitations. 
First of all, we model edge weights as categories, 
which works well for binned edge weights in practice but is theoretically dissatisfying. 
Therefore, a natural enhancement of \ourmethod would be able to handle real edge weights, 
possibly using a maximum entropy model on its edge weight distribution.
Second, we currently test all our graph patterns at the same alpha level. 
While this is theoretically defensible, given that we combine our statistical test with a model selection criterion, 
dynamically adjusting our alpha level might be an option worth exploring.
Finally, \ourmethod is currently limited to groups of node-aligned graphs, 
and extending it to other graph types constitutes an open opportunity for future work.

\cleardoublepage

\bibliography{bib/abbrev,bib/bib-jilles,bib/bib-paper}

\end{document}